\documentstyle[aps,prl]{revtex}
\begin{document}
\bibliographystyle{unsrt}
\title{Theory of 2D metal-insulator transition I: Zero Magnetic field}
\author{Konstantin Kladko}
\address{Sun Microsystems,  MS- CUP02-302, \\
901  San Antonio Road, Palo Alto,  94303 CA United States\\
e-mail:Konstantin.Kladko@sun.com}
\maketitle

\begin{abstract}
We propose a scaling theory of 2D metal insulator transition discovered by
Kravchenko and coworkers. In this theory conductance/resistance duality is
an exact relation. The exponent of the stretched exponential in $\sigma(T)$
is determined by the temperature dependence of the coherence length.
The quantum-chaotic nature of the transition is manifested in the 
square form of the phenomenological scaling function.
\end{abstract}

The 2D metal insulator transition (MIT) discovered by Kravchenko and
coworkers is  subject of wide interest. Nevertheless, to our knowledge,
no satisfactory theory of such a transition has been proposed.
Moreover, the discovery of such a transition puts in a serious question
(in our opinion disproves) the validity of previous theories in this field.
Here we propose a phenomenological scaling theory, which accounts for 
the following properties of the MIT at zero magnetic field. All
of these properties were confirmed in different experiments on 
different systems. In our view, these properties summarize all
interesting facts, measured in transport so far.

1. Resistance and conductance have a dual symmetry with respect to the
critical point  $\sigma =\sigma _0$. 

2. At the critical point  conductance does not depend on temperature. 

3. All $\sigma (T)$ curves scale to the same universal curve.

4. $\sigma (T)$ shows approximate stretched exponential behavior $\simeq
\exp{ CT^x}$. The value
of the stretched exponential $x$ is not universal, and is different in different
experiments.

To formulate a scaling theory, in the spirit of Landau, we need to guess the simplest
phenomenological scaling relation for the conductance of the sample. It is clear,
that at sample lengths larger, than the coherence length, quantum interference
effects do not exist. Therefore, at such distances quantum scaling stops and
conductance satisfies classical conductance addition laws. Let us imagine a square sample 
of length $L$, which has conductance $\sigma (L)$. Now we need to find a scaling equation
which relates $\sigma (L)$ to  $\sigma (2L)$, such as $\sigma(2L) = F(\sigma(L))$.
The function $F(\sigma)$ is the main ingredient of the phenomenological theory.
This function should be postulated by the comparison to the experiment. 
There is no need to derive it from anything else, since the underlying strongly
interacting electron system is inherently chaotic and self-organized.
We could, therefore, postulate $F(\sigma)$ without any derivations. We
will, though, give a qualitative explanation for the chosen form of $F(\sigma)$.

In Landauer formalism (Fig.1)
we phenomenologically relate $\sigma(L)$ to the number of conducting
scattering channels, which go from the left side of the sample to the right
side. Now let us consider squares $A$ and $B$ in Fig. 1. The conductance 
$\sigma (2L)$ will be then proportional to the probability for two $L-$%
channels, one from $A$ and one from $B,$ to meet and to form a long $2L$%
-channel penetrating the larger square. Assuming, that excited state wave
functions are very complicated, and channels meet chaotically, we can
conclude, that the number of these $2L$-channels will be proportional to the
squared density of $L$-channels, i.e. to the squared conductance $\sigma (L)$%
. These arguments, therefore, lead to the following scaling equation
\begin{equation}
\sigma (2L)=C\sigma ^2(L)
\end{equation}
The function $F(\sigma)$ is, therefore, a quadratic function. We reiterate, that
the above derivation is only a plausible analogy, which does not make an attempt
to fully explain the complex structure of chaotic many-particle wave functions.
It merely illustrates, that the quadratic $F(\sigma)$ is a right choice for the
phenomenological theory.
At the critical conductance resistance should be constant. Therefore, $C=1/$ 
$\sigma _0$, and

\begin{equation}
\sigma (2L)=\sigma ^2(L)/\sigma _0
\end{equation}

Now, to find $\sigma (T)$ we need to integrate this equation from the
microscopic length $l$ to the coherence length $L_{coh}(T)$. Assuming,
that the conductance at the microscopic length scale $l$ is $\sigma_l$, 
the solution is
\begin{equation}
\sigma (T)=\sigma_0 (\sigma_l/\sigma _0)^{L_{coh}(T)/l}
\end{equation}

Here $\sigma_l$ is the conductance at the microscopic length scale $%
l$. $\sigma_l$  depends on carrier density. From here on we will measure
$\sigma$ in units of $\sigma_0$. Then the above formula reads

\begin{equation}
\sigma (T)=\sigma_l^{L_{coh}(T)/l}
\end{equation}

The above
formula has a duality property, namely, transformation $\sigma_l
-> 1/\sigma_l$ sends conductance into resistance and
vice versa. This is, in our view, the duality which was observed in the
experiment. If we assume, that $\sigma_l$ smoothly depends on density 
near the transition, then near the transition one can always find such
densities $n$ and $n'$, that they will be in a duality relation. 
We note, that the duality relation is exact and does not depend 
on the specific temperature dependence of the coherence length. 
Now, if the coherence length  goes as a power law of temperature $L_{coh}=A/T^\alpha$, that one usually assumes, one obtains  a stretched 
exponential dependence of $\sigma (T)$. 

\begin{equation}
\sigma (T)= \sigma_l^{\frac A{lT^\alpha }}=e^{\frac A{lT^\alpha }\ln
(\sigma_l)}
\end{equation}

Let us note that this dependence
is approximate. Moreover, different stretched exponents observed in different
experiments should correspond to different behaviors of the coherence 
length as a function of temperature in different systems. This resolves
the question, why duality is universal and the value of the stretched exponential
is, probably not.
Nonuniversal values of $\alpha $ found in different experiments could mean,
that the coherence length is set by external to the transition factors
(phonons etc.). 

We also, note, that in order for conducting phase to appear, $\sigma_l$,
measured in units of $\sigma_0$ should become larger, than $1$. The
reason, why conducting phase never shows in lower mobility samples,
is probably because $\sigma_l$ there never gets large enough.

To incorporate magnetic field, one needs to include a phenomenological
magnetic-field-dependent term into the right-hand side of the scaling
equation. This will be done elsewhere. We believe, that this can be done
consistently. We also believe, that phenomenological approach is the only 
approach which is meaningful in the case of the 2D MIT transition, which
has inherently collective and chaotic nature.

I thank Sergej Kravchenko, Yaroslav Bazaliy and Pavel Kornilovich for 
enlightening comments and suggestions.  This work has been supported by the 
Otto Hahn Medal Fellowship of the Max Planck Society, Germany.
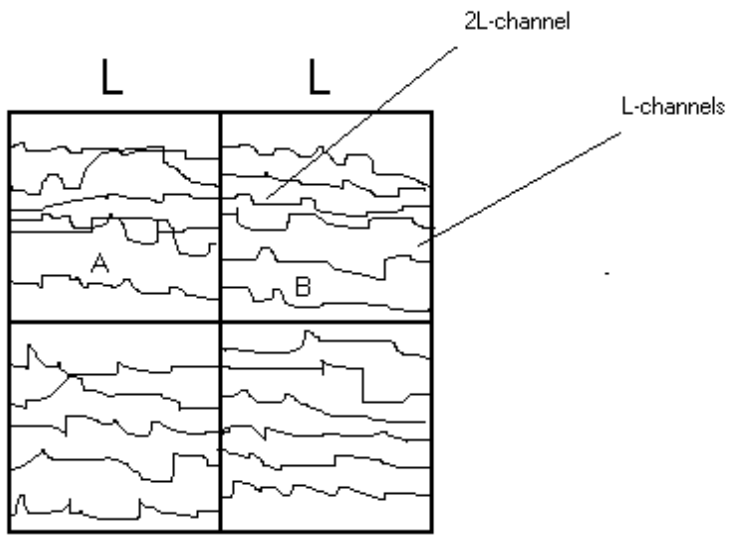
\end{document}